\documentclass[]{aastex631}

\submitjournal{ApJ}

\usepackage{color}
\definecolor{bbsalmon}{rgb}{1.0, 0.47, 0.42}

\begin{document}

\title{Constraining Supernova Physics through Gravitational-Wave Observations}

\correspondingauthor{Gergely D\'{a}lya}
\email{gergely.dalya@ugent.be}

\author[0000-0003-3258-5763]{Gergely D\'{a}lya}
\affiliation{Department of Physics and Astronomy, Universiteit Gent \\
B-9000 Ghent, Belgium}

\author[0000-0002-2566-4888]{Sibe Bleuz\'{e}}
\affiliation{Department of Physics and Astronomy, Universiteit Gent \\
B-9000 Ghent, Belgium}

\author[0000-0003-0909-5563]{Bence B\'{e}csy}
\affiliation{eXtreme Gravity Institute, Department of Physics, Montana State University\\
Bozeman, Montana 59717, USA}
\affiliation{Department of Physics, Oregon State University, Corvallis, OR 97331, USA}

\author[0000-0001-7207-4584]{Rafael S. de Souza}
\affiliation{Key Laboratory for Research in Galaxies and Cosmology, Shanghai Astronomical Observatory\\ Chinese Academy of Sciences, 80 Nandan Rd., Shanghai 200030, China}

\author[0000-0003-4610-1117]{Tam\'{a}s Szalai}
\affiliation{Department of Experimental Physics, Institute of Physics, University of Szeged, H-6720 Szeged, Dóm tér 9, Hungary}
\affiliation{ELKH-SZTE Stellar Astrophysics Research Group, H-6500 Baja, Szegedi {\'u}t, Kt. 766, Hungary}



\begin{abstract}

We examine the potential for using the LIGO-Virgo-KAGRA network of gravitational-wave detectors to provide constraints on the physical properties of core-collapse supernovae through the observation of their gravitational radiation. We use waveforms generated by 14 of the latest 3D hydrodynamic core-collapse supernova simulations, which are added to noise samples based on the predicted sensitivities of the GW detectors during the O5 observing run. Then we use the BayesWave algorithm to model-independently reconstruct the gravitational-wave waveforms, which are used as input for various machine learning algorithms. Our results demonstrate how these algorithms perform in terms of i) indicating the presence of specific features of the progenitor or the explosion, ii) predicting the explosion mechanism, and iii) estimating the mass and angular velocity of the progenitor, as a function of the signal-to-noise ratio of the observed supernova signal. The conclusions of our study highlight the potential for GW observations to complement electromagnetic detections of supernovae by providing unique information about the exact explosion mechanism and the dynamics of the collapse.

\end{abstract}

\keywords{Core-collapse supernovae (304) --- Gravitational waves (678) --- LIGO (920)}


\section{Introduction} \label{sec:intro}

The global network of gravitational-wave detectors, including Advanced LIGO~\citep{Aasietal2015}, Advanced Virgo~\citep{Acernese_2014} and KAGRA~\citep{akutsu2020overview}, has already observed 90 gravitational wave (GW) candidates originating from the coalescence of compact astrophysical objects \citep{2021arXiv210801045T,2021arXiv211103606T}. Besides compact binary coalescences, other sources can also generate GWs potentially observable with the current detector network. One of these sources are core-collapse supernovae (CCSNe), which are extremely powerful explosions at the last evolutionary stage of stars with masses exceeding $\simeq 8\ M_{\odot}$, corresponding to $\sim 70$\% of all supernovae in the local Universe \citep{li11}. There are both optically targeted \citep{2020PhRvD.101h4002A} and general all-sky, all-time transient searches \citep{2021PhRvD.104l2004A} performed by the LIGO-Virgo-KAGRA Collaboration (LVK) looking for GW signals emitted by CCSNe. Still, so far, none of these has achieved significant detection.

CCSNe are among the most promising candidates for a multi-messenger observation through the detection of electromagnetic radiation, GWs and neutrinos from the same source \citep{1987PhRvL..58.1490H}. GWs and neutrinos could provide unique information about the exact mechanism of the explosion, as well as the dynamics of the collapse, which could complement the information gained through electromagnetic observations, as the latter emission is delayed by a timescale of hours and originates in regions farther away from the central engine \citep{2013ApJ...778...81K}. 

SN theory and simulations made huge progress in the last decade, leading to a better understanding of the physical mechanisms occurring during CCSN explosions \citep{2021Natur.589...29B}. Once the mass of the iron core of a star exceeds the Chandrasekhar mass, the core cannot be stabilized by the electron degeneracy pressure anymore and it starts to collapse. The collapse stops and the matter of the core, called the proto-neutron star bounces back when the density reaches the nuclear density. The bounce initiates a shock wave moving outwards, however, in itself the shock wave would not have enough energy to reach the surface of the star leading to the explosion, so it stops after a few tens of milliseconds and needs another process to restart it \citep{2012ARNPS..62..407J}. 

One of the possible shock revival mechanisms is the delayed neutrino-heating mechanism, proposed by \citet{1985ApJ...295...14B}. Around $\sim$100~ms after the stalling of the shock wave, the conditions become ideal for an efficient energy deposition from neutrinos to the postshock medium. If the amount of energy transferred is sufficient, the rising pressure can accelerate the shock outwards against the ram pressure of the collapsing star's outer layers. Hence, in this explosion mechanism, the expansion of the shock is driven by the energy deposition of the neutrinos \citep{Janka2017}.

For rapidly rotating stars having strong magnetic fields, magnetohydrodynamic phenomena, particularly the magneto-rotational mechanism, proposed by \citet{1971SvA....14..652B} can initiate the shock revival. With strong enough magnetic fields and high enough angular velocity of the rotation, it is possible to revive the shock wave by the transfer of energy from the highly magnetized proto-neutron star outward to the stalled shock wave, pushing the material towards the surface (see e.g.~\citealt{2020MNRAS.492.4613O}). Note, that this process is only expected to account for $\sim 1\%$ of all CCSN explosions \citep{2006ApJ...637..914W}.

Simulations show a number of interesting phenomena occurring under the stellar surface during the imploding and exploding phases, like sloshing movements, convection and oscillatory behaviour. GWs can give us a unique view to explore these phenomena, which leave their own unique imprint on the GW signal emitted. In this study we investigated the possibility of detecting two of these events:
\begin{itemize}
    \item Prompt postbounce convection: When the supernova shock wave stalls, there are strong gradients left in its path, since it blew almost everything away in front of itself. An example is the negative lepton gradient, caused by the increased interaction of electron neutrinos resulting in a drop in their density. These gradients drive a prompt convection in front of and behind the shock wave \citep{1996A&A...306..167J}. This in itself may then cause even more energy-producing interactions, thus creating another way of refueling the shock wave. While its exact impact may be still up for discussion, prompt convection itself has been observed in simulations, so determining its presence in observations would certainly matter in establishing the different contributions to shock revival.
    \item Standing accretion shock instability (SASI): While carrying out idealized 2D simulations of an accretion shock, \citet{Blondin_2003} found that a standing accretion shock in an axisymmetric environment would become unstable under radial perturbations. The SASI emerges as asymmetric sound waves characterized by low-order spherical harmonics ($\ell = 1, 2$). These waves grow until they eventually begin to impact the shock wave itself. 
    
    One effect of the SASI is the expansion of the average shock radius. Its sloshing-type motion thus generally pushes material outwards, helping to revive the stalled shock front. This makes it potentially crucial in the explosion mechanism, hence its imprint should be searched for in GW observations. A second reason is its potential connection to the equation of state. While it was discovered in 2D simulations, a lot of 3D simulations also show SASI activity. The difference is that in 3D it has more complex, nonaxisymmetric modes and has been observed to redistribute angular momentum.
\end{itemize}

The field of constraining different areas of SN physics using GW observations is an active area of research with many interesting recent results (see e.g.~\citealt{2014PhRvD..90l4026E}, \citealt{2014PhRvD..90d4001A}, and \citealt{2019PhRvD..99f3018R}). \cite{2022PhRvD.105f3018P} performed a study using an asymmetric chirplet signal model and four different hydrodynamical SN simulations to show how we might place constraints on the mass and radius of the proto-neutron star (PNS), among other parameters. \cite{2021PhRvD.103f3006B} initiated a programme to infer the properties of the PNS based on the gravitational waves of its oscillations, and in their recent paper \citep{2023arXiv230110019B} they extend the previous work by coherently combining the data from several GW detectors and by inferring the time evolution of a combination of the mass and radius of the PNS.

In this paper we examine how stringent constraints could we give on the physical properties of a CCSN through the observation of its gravitational radiation with the LIGO-Virgo-KAGRA network. We use GW waveforms generated by 14 of the latest 3D hydrodynamic CCSN simulations. The waveforms are injected into noise samples derived from the predicted sensitivities of the GW detectors during the O5 observing run \citep{2018LRR....21....3A}, and reconstructed with the BayesWave algorithm before used as the input of different machine learning (ML) algorithms. We present how the ML algorithms trained with this sample perform as a function of the signal-to-noise ratio (SNR) of the observed CCSN signal, by means of their accuracy of:
\begin{itemize}
    \item indicating the presence of specific features of the progenitor or the explosion (i.e. rotation, prompt convection, and SASI),
    \item predicting the shock revival mechanism (i.e. magneto-rotational or neutrino-driven),
    \item estimating the mass and the angular velocity of the progenitor.
\end{itemize} 
Using ML algorithms to detect and classify GW signals originating from SNe is an emerging field, see e.g.~\cite{2021PhRvD.103j3008M}, \cite{2022PhRvD.105h4054A}, and \cite{2023A&A...669A..42I}.

This paper is organized as follows. In Section \ref{sec:methods}, we describe the methods we used for generating simulated signals and reconstructing those with BayesWave, followed by introducing the ML techniques used for this analysis. In Section \ref{sec:results}, we present the results of our analyses regarding the performance of the different ML algorithms. We summarize our findings and draw conclusions in Section \ref{sec:conclusion}.



\section{Methods} \label{sec:methods}

In this section we give an overview of the various tools we used, starting with the CCSN models we used to generate GW signals (Section \ref{subsec:SN-waveforms}). The signals were reconstructed using BayesWave, the exact process of which is discussed in Section \ref{subsec:BayesWave}. We used various dimensionality reduction methods on the reconstructed GW signals and used those to train and test classification and regression algorithms (see Section \ref{subsec:ML}). 

\subsection{Supernova waveform models}
\label{subsec:SN-waveforms}

In order to test what physical constraints could be given to supernovae after a successful GW detection, we used 14 different waveform families, all originating from 3D hydrodynamical simulations. We have chosen most of them from among the waveform families used by \citet{2021PhRvD.104j2002S}, who performed a detailed study on the detectability of GWs from various SN simulations. We have extended the list of 3D waveform families by several newer models which became public after the publication of \citet{2021PhRvD.104j2002S}. The waveform families we used for this simulation are as follows:

\begin{itemize}
    \item S11 \citep{10.1093/mnras/stx618}\\
    The S11 waveform family is the one with the lightest progenitor that is discussed in \citet{10.1093/mnras/stx618}. Contrary to its $20~M_{\odot}$ and $27~M_{\odot}$ counterparts from the same study, this $11.2~M_{\odot}$ zero-age main sequence (ZAMS) mass model does not show strong SASI activity or prompt convection. Non-resonant g-modes are present, whereas resonant g-mode oscillation is suppressed compared to previous (2D) models. The supernovae simulated by this waveform family are neutrino-driven. Over the course of 350~ms after core bounce, they produce GWs with an energy of roughly $1.1\cdot10^{-10}~M_{\odot} c^2$ with a peak frequency around 642~Hz.
    
    \item M15FR \citep{10.1093/mnras/stz990}\\
    \citet{10.1093/mnras/stz990} simulated three $15~M_{\odot}$ ZAMS mass models, of which the fast rotating one (M15FR) is used in this study. With an artificially enhanced angular velocity of $0.5~\mathrm{rad}\,\mathrm{s}^{-1}$ comes also powerful SASI activity, resulting in a higher GW energy than the previous model at $2.7\cdot10^{-10}~M_{\odot} c^2$. The neutrino-driven explosion reaches a peak frequency around 689~Hz during the 460~ms that were simulated after core bounce. One of the effects of the fast rotation is that unlike in the slowly rotating and non-rotating models, resonant g-modes are present in this model.
    
    \item H \citep{10.1093/mnras/stab2161}\\
    To study the explosion dynamics of SNe with complex magnetic structures, \citet{10.1093/mnras/stab2161} simulated SN explosions of fast rotating $35~M_{\odot}$ ZAMS mass stars with various magnetic configurations. From these, we have chosen the hydrodynamic benchmark H model with no magnetic fields, leading to the weakest emissions and the most massive and rapidly rotating PNS in their simulations due to the lack of efficient magnetic extraction of rotational energy \citep{2022arXiv221005012B}. 
    
    \item TM1 \citep{2016ApJ...829L..14K}\\
    While having the same $15~M_{\odot}$ ZAMS mass as the M15FR model, this waveform family is non-rotating and uses a different equation of state (EoS). \citet{2016ApJ...829L..14K} studied the effect of different equations of state on the supernova characteristics and found that the softer the EoS is, the more the SASI develops. This means that the presence and strength of SASI activity holds information about the EoS as well, which is one of the reasons detecting SASI activity would be quite exciting. Besides SASI activity, g-mode oscillation is also present in the neutrino-driven explosion simulated here. The model reaches a total GW energy of $1.7\cdot10^{-9}~M_{\odot} c^2$ in the first 350~ms post bounce and has a peak frequency of 714~Hz.
    
    \item S11.2 \citep{2017ApJ...851...62K}\\
    This $11.2~M_{\odot}$ ZAMS mass model has a softer EoS than S11 from \citet{10.1093/mnras/stx618}, which leads to a noticeable presence of SASI. This non-rotating neutrino-driven model also develops prompt convection. In the simulated 190~ms post bounce the peak frequency is 195~Hz and the emitted GW energy is $1.3\cdot10^{-10}~M_{\odot} c^2$.
    
    \item C15-3D \citep{PhysRevD.102.023027}\\
    Similarly to the model TM1, this model is also a non-rotating, $15~M_{\odot}$, neutrino-driven model exhibiting SASI and g-mode activity. However, its peak frequency of 1064~Hz is relatively high compared to similar models and the bandwidth covered by the waveform is also significantly more spread out in the simulation lasting 420~ms.
    
    \item Mesa20\_pert \citep{2018ApJ...865...81O}\\
    This simulation started from a non-rotating $20~M_{\odot}$ progenitor. 
    Like several other models used, it is also a neutrino-driven explosion with SASI and g-modes present. Its peak frequency is 1033~Hz and it produces GWs with an energy of $1.7\cdot10^{-9}~M_{\odot} c^2$ in the 530~ms after the core bounce.
    
    \item S27-fheat1.00 \citep{2013ApJ...768..115O}\\
    The progenitor of this model is a $27~M_{\odot}$ star. The non-rotating, neutrino-driven explosion produces $4\cdot 10^{-10}~M_{\odot} c^2$ energy in 190~ms post bounce. Prompt convection and SASI are present in the waveform.
    
    \item S40\_FR \citep{2021ApJ...914..140P}\\
    This is the model with the heaviest progenitor ($40~M_{\odot}$) among the rapidly rotating ones we included in this study. The angular velocity is $1~\mathrm{rad/s}$, and there is SASI and prompt convection present in the waveform produced by a neutrino-driven explosion.
    
    \item S18 \citep{10.1093/mnras/stz1304}\\
    This neutrino-driven explosion shows the excitation of surface g-modes, but lacks emission due to SASI activity. The GWs emitted by this non-rotating source having a progenitor mass of $18~M_{\odot}$ reach a total energy of $1.6\cdot10^{-8}~M_{\odot} c^2$.
    
    \item M39 \citep{10.1093/mnras/staa1048}\\
    This $39~M_{\odot}$ model is a rapid rotator ($0.542~\mathrm{rad/s}$), showing both SASI activity and prompt convection. One of the important findings of \cite{10.1093/mnras/staa1048} is that despite having a neutrino-driven explosion, the rotation and SASI aid in producing GWs which they claim to be strong enough to be detectable out to 2~Mpc with the future Einstein Telescope.
    
    \item Z100\_SFHx \citep{10.1093/mnras/stab614}\\
    The heaviest of all the models we use in this study is the one developed by \cite{10.1093/mnras/stab614}. The non-rotating source has a progenitor mass of $100~M_{\odot}$. The neutrino-driven explosion develops strong SASI activity.
    
    \item S9 \citep{2019ApJ...876L...9R}\\
    This non-rotating model has the smallest progenitor mass ($9~M_{\odot}$) of all the waveform families considered here. With a few low-mass, neutrino-driven models, \cite{2019ApJ...876L...9R} confirm that for a detection to happen at the next galactic supernova event, next-generation GW detectors will be needed for stars with relatively low masses. With a simulation extending to 1100~ms post bounce, it reaches a GW energy of $1.6\cdot10^{-10}~M_{\odot} c^2$. SASI activity is not visible in the GW signal, but prompt convection is.
    
    \item R4E1FC\_L \citep{2010A&A...514A..51S}\\
    This is the second MHD-driven model in our selection. With an angular velocity of $9.4~\mathrm{rad/s}$, this model has a GW energy output of $3.9\cdot10^{-7}~M_{\odot} c^2$. During the rather short simulation (100 ms), the authors identify the presence of prompt convection.
    
\end{itemize}

\subsection{BayesWave}
\label{subsec:BayesWave}

BayesWave \citep{BayesWaveI,BayesWaveII,BayesWaveIII} is a Bayesian analysis pipeline providing model-agnostic waveform reconstruction for gravitational-wave signals. It does so by modeling the signal as a sum of Morlet-Gabor wavelets while marginalizing over the number of wavelets using a trans-dimensional sampler. BayesWave has been extensively tested and used to analyze a wide variety of GW signals including ad-hoc waveforms \citep{BayesWavePE}, circular black hole binaries \citep{BayesWavePE}, eccentric black hole binaries \citep{BayesWaveEcc}, neutron star post-merger signals \citep{BayesWaveNSPostMerger}, black hole post-merger echoes \citep{BayesWavePostMergerEcho}, and most importantly core-collapse supernova signals \citep{2018arXiv180207255G,BayesWaveSN}. As BayesWave is one of the widely used waveform reconstruction tools in the LVK we expect that in the case of a real CCSN detection it will also be among the pipelines used by the LVK for analysing the data.

\cite{BayesWaveSN} optimized the settings of BayesWave for CCSN detections, and in this study we followed the steps outlined in the paper to maximize our reconstruction accuracy, namely, by increasing the allowed wavelet quality factor to 100, increasing the number of iterations in the reversible-jump Markov chain Monte Carlo to 4 million, setting the lower end of the allowed frequency range to 32 Hz and the sampling rate to 4 kHz. Note, that here we did not exploit the multi-messenger nature of CCSNe, i.e.~that the sky location could be known from the EM counterpart. That would allow BayesWave to recover the waveforms more accurately, but we only relied on the GW signal, giving conservative estimates for a real detection. Using wavelets with a frequency evolution, called 'chirplets' in BayesWave can lead to more accurate waveform reconstructions in several cases (see e.g.~\citealt{BayesWaveEcc}), however as \cite{BayesWaveSN} has shown that it is not necessarily the case for CCSN signals, we did not use chirplets for this study.

\subsection{Machine learning}
\label{subsec:ML}

The BayesWave algorithm provides various forms of output data, including a time-domain waveform, a power spectrum, and a spectrogram. Our study uses the power spectra, as the time domain representation may introduce bias in the machine learning algorithms. We applied dimensionality reduction methods to the BayesWave reconstructions, and then used these as input for the training and testing of various classification and regression techniques. The properties of the algorithms we employed are summarised in the following sections.

\subsubsection{Dimensionality reduction}

Dimensionality reduction is a crucial step in analysing high-dimensional data. Linear methods such as Principal Component Analysis (PCA) and Linear Discriminant Analysis (LDA) are commonly used, as they are computationally efficient and easy to interpret. PCA is a widely used method for reducing the dimensionality of data and exploring its underlying structure. By performing eigendecomposition or singular value decomposition on the input data, PCA generates a new set of orthogonal variables, known as principal components, that capture as much of the variability in the original data as possible. This technique has been extensively applied in various fields, including astronomy. It has been used to analyse exoplanet imaging data, classify Type Ia supernovae, separate foregrounds in 21 cm intensity maps, and reconstruct point spread functions.
For datasets with complex and non-linear relationships, non-linear methods such as t-SNE \citep{tsne}, UMAP \citep{McInnes2018,2018arXiv180203426M} and Isomap \citep{isomap} are more suitable. 
These methods use more complex mathematical functions to project the data into a lower-dimensional space while preserving the non-linear structure of the data.

We tested both PCA and UMAP as dimensionality reduction algorithms on the power spectra reconstructed by BayesWave. We performed a couple of preliminary tests to find a suitable number of principal components. We found that decreasing the number of components such that they explain 98\% of the variance (which meant a factor of $\sim50$ reduction in the number of components) resulted in more accurate results of the ML algorithms than keeping more or less components, however, there might be room for further optimization here in the future. We used the same number of components for both PCA and UMAP. Following that, we used 90\% of the data as the training sample to train the various classification and regression algorithms, and the remaining 10\% of the data to test their performance.


\subsubsection{Classification and regression}

\paragraph{Classification}  A supervised learning task that involves predicting a discrete label or class for a given input. This paper considers three popular classification methods: decision tree (DT), support vector machine (SVM), and k-nearest neighbour (k-NN). Below we summarise their pros and cons.

\begin{itemize}

\item A decision tree is a tree-based model that recursively partitions the input space into smaller regions, each associated with a class label. The decision tree algorithm constructs the tree by choosing the feature and threshold that maximises the separation between different classes at each step. Decision trees are easy to interpret and can handle numerical and categorical features. However, they are prone to over-fitting when the tree becomes too deep.

\item  SVM is a linear model that finds the hyperplane that maximally separates the different classes in the feature space. SVM can handle non-linearly separable data by introducing a kernel trick that maps the input space into a higher-dimensional feature space.
SVM is a robust and powerful method, but it can be sensitive to the choice of kernel and parameters.

\item k-NN is a non-parametric method that assigns a class label to a given input based on the majority vote of its k-nearest neighbours in the training set.
k-NN is simple and easy to implement, but it can be computationally expensive and sensitive to the choice of k.
\end{itemize}

\paragraph{Regression} It involves predicting a continuous output value for a given input. This paper considers two popular regression methods: linear regression and LASSO.

\begin{itemize}

\item Linear regression is a linear model that finds the coefficients of a linear equation that best fits the training data. The method assumes a linear and additive relationship between the input and output variables. 
It is a simple and interpretable method but can be sensitive to outliers and irrelevant features.

\item LASSO (Least Absolute Shrinkage and Selection Operator) is a regularisation method that introduces a penalty term on the absolute values of the coefficients in the linear regression equation. LASSO can select a subset of relevant features by shrinking the coefficients of irrelevant features to zero. LASSO is a powerful method for feature selection, but it can be sensitive to the choice of the regularisation parameter.
\end{itemize}

\section{Results} \label{sec:results}

The inference of several characteristics of the various models is a classification problem, i.e.~the given property has distinct classes. These include the explosion mechanism, the presence of SASI, prompt convection and rotation in the model. To characterize the efficiency of these classifications we used the balanced accuracy as our evaluation metric. It is calculated as the mean of the true positive rate and the true negative rate:
\begin{equation}
    \mathrm{Balanced\ accuracy} = \frac{1}{2}\left(\frac{\mathrm{TP}}{\mathrm{TP}+\mathrm{FN}} + \frac{\mathrm{TN}}{\mathrm{TN}+\mathrm{FP}}\right),
\end{equation}
where TP, TN, FP and FN are the number of true positives, true negatives, false positives and false negatives, respectively. The balanced accuracy becomes equivalent to the traditional accuracy if the classifier performs equally well on both classes. This metric is particularly useful when the dataset is imbalanced, as it accounts for potential bias towards the majority class.

\begin{figure*}
    \centering
    \includegraphics[width=.95\columnwidth]{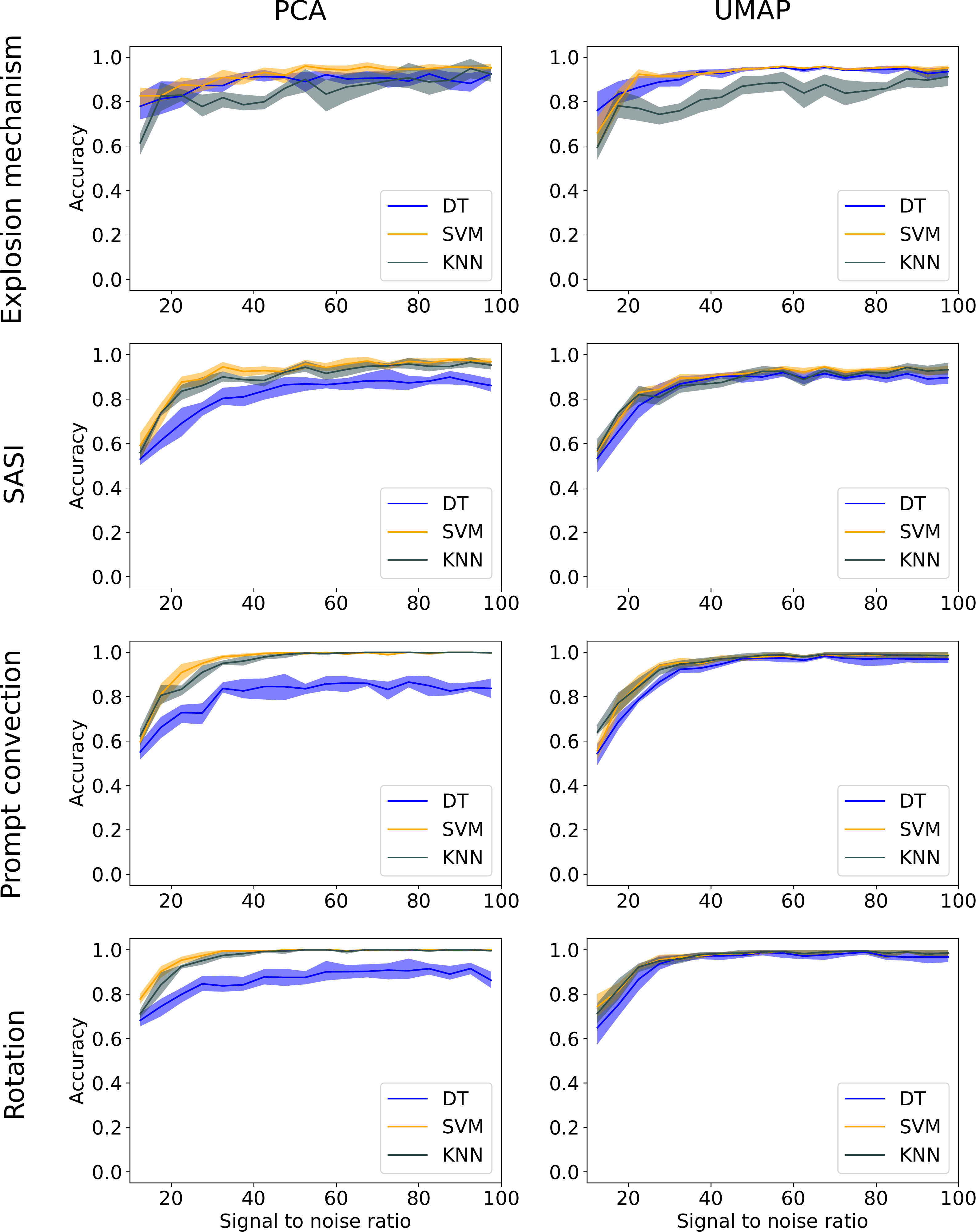}
    \caption{Balanced accuracy for the predicted explosion mechanism (first row), the presence of SASI (second row), the presence of prompt convection (third row), and the presence of rotation (fourth row) as a function of SNR for both dimensionality reduction techniques. The left column shows the results we got using PCA, while the right column shows the results using UMAP. The colors of the curves indicate the classifier used: blue for decision tree, orange for support vector machine and green for k-nearest neighbour. Shaded areas represent the $1\sigma$ uncertainties obtained over 10 realizations.}
    \label{fig:classification}
\end{figure*}

Figure \ref{fig:classification} shows the results we got for the classification problems using both PCA (left column) and UMAP (right column) for dimensionality reduction, and DT, SVM and KNN as classifier. For each parameter and for each dimensionality reduction technique, the figure shows the average performance of the predictor. The shaded areas correspond to the standard deviation on this average, calculated on 10 distinct executions of the algorithms. The samples were binned in SNR bins of width 5 before calculating the error. As expected, with higher SNR we get a higher accuracy in the case of all features, using both reduction techniques and all classifiers. For nearly all cases we get a $>90\%$ balanced accuracy above $\mathrm{SNR}\gtrsim 30$. A notable exception seems to be the combination of PCA+DT, where the accuracy saturates to $\sim 0.9$. The underperformance of UMAP+KNN compared to the other combinations when classifying the explosion mechanism was caused mainly by a confusion between the Andresen M15FR and the Bugli H models; these two waveforms do not have as many distinctive features as the other ones, which led to them having similar parametrizations in the reduced parameter space.

The two regression problems we studied were the inference of the progenitor mass and the angular velocity. We characterize the efficiency of the regressions using the normalized root mean squared error (NRMSE), which can be calculated as the ratio of the root mean squared error of the model to the standard deviation of the observed data:
\begin{equation}
    \mathrm{NRMSE} = \frac{1}{\sigma_{\mathrm{O}}} \sqrt{\frac{1}{n}\sum_{i=0}^{n}(y_{i,\mathrm{O}}-y_{i,\mathrm{P}})^2},
\end{equation}
with the differences involving the original (O) and the predicted (P) set of parameters and $n$ samples in the test set with a standard deviation of $\sigma_{\mathrm{O}}$. This metric provides a way to compare the error of a model to the variability of the data, allowing for a more meaningful interpretation of the model's performance.

\begin{figure*}
    \centering
    \includegraphics[width=.95\columnwidth]{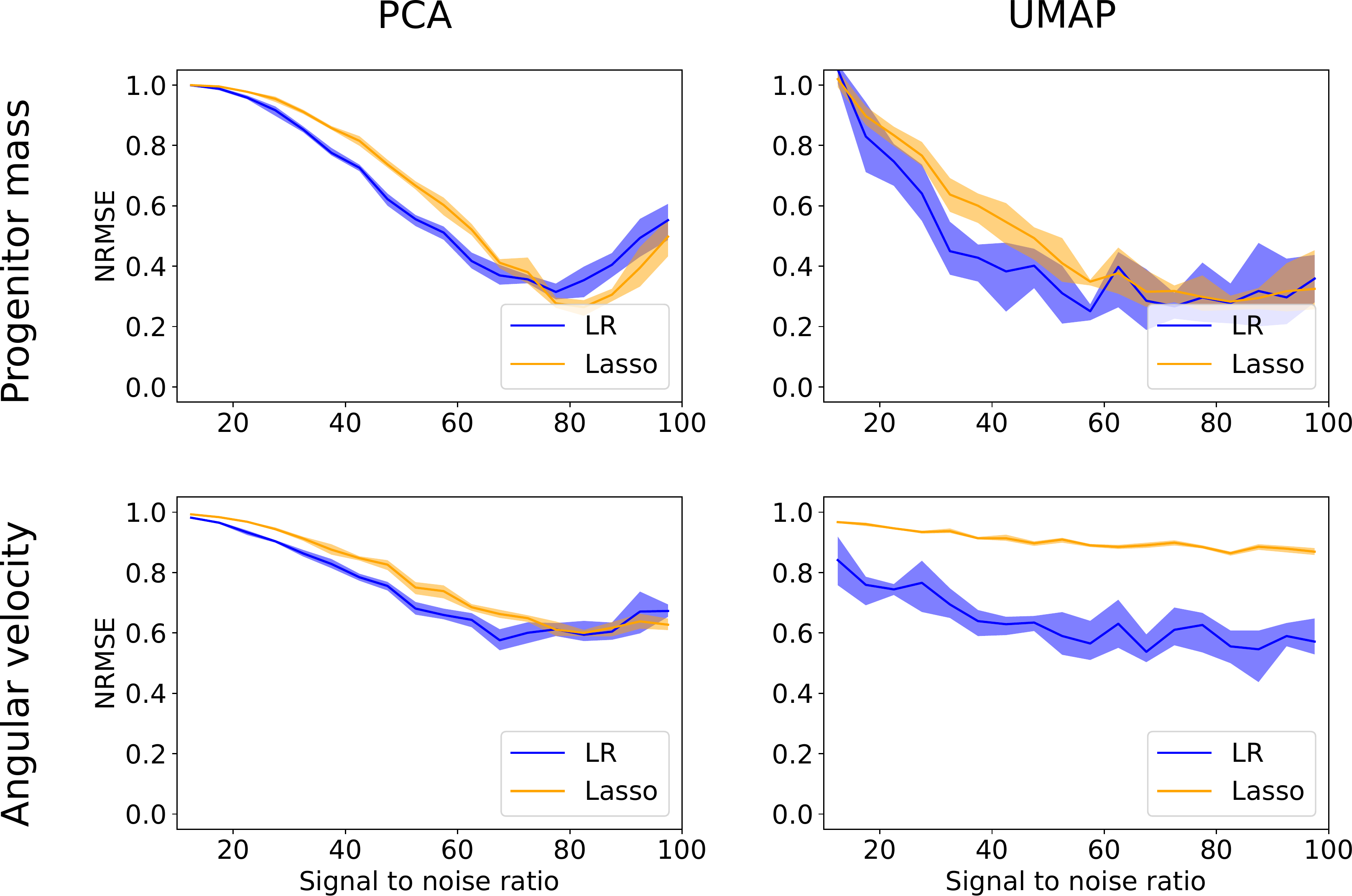}
    \caption{Normalized root mean squared error on the predicted progenitor mass (top row) and on the predicted angular velocity (bottom row) as a function of SNR for both dimensionality reduction techniques. The left column shows the results we got using PCA, while the right column shows the results using UMAP. The colors of the curves indicate the regression method used: blue for linear regression and orange for LASSO. Shaded areas represent the $1\sigma$ uncertainties obtained over 10 realizations.}
    \label{fig:regression}
\end{figure*}

Figure \ref{fig:regression} shows the NRMSE on the predicted progenitor masses (top row) and angular velocities (bottom row) as a function of SNR using both PCA (left column) and UMAP (right column) as dimensionality reduction techniques. As expected, the error decreases for higher SNR values, however, using PCA we see an increasing error above $\mathrm{SNR}\gtrsim 80$. For most of our cases, the PCA-based predictors show a more stable behaviour, while the UMAP-based ones show more variance in the error. Generally, the predictions of progenitor mass are more accurate than those of the angular velocity. The reason behind this is that the masses of the models used sampled the whole range of possible progenitor masses more evenly and there were also several SN models having similar progenitor masses but differing in other physical parameters. However, most of our models were non-rotating, so the distinct angular velocity values of the 5 rotating models were not enough to cover the parameter space properly and enable us to distinguish the effect of the different angular velocity from the effect of the other physical parameters. The predictions for both the progenitor mass and the angular velocity reach a plateau at the high-SNR limit, which shows that systematic modelling uncertainties start to dominate over the statistical errors of the observations. At this regime, the precision of the estimations can only be increased by having more simulations with different progenitor masses and angular velocities.

\subsection{Using an unknown model}

\begin{figure*}
    \centering
    \includegraphics[width=.95\columnwidth]{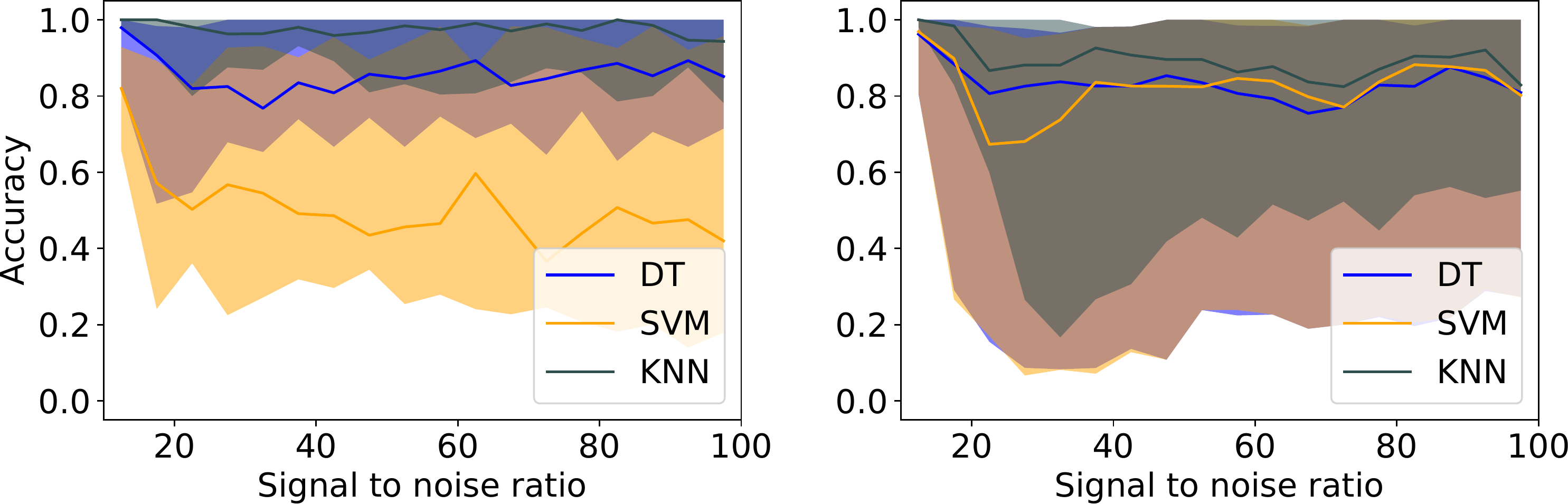}
    \caption{Balanced accuracy for the predicted shock revival mechanism as a function of SNR for both dimensionality reduction techniques. The left column shows the results we got using PCA, while the right column shows the results using UMAP. The colors of the curves indicate the classifier used: blue for decision tree, orange for support vector machine and green for k-nearest neighbour. Shaded areas represent the $1\sigma$ uncertainties. The plots show the averaged result we got by leaving out each of the neutrino-driven models one by one from the teaching sample and trying to predict the shock revival mechanism of the left-out model with the algorithms set up this way.}
    \label{fig:leftout}
\end{figure*}

In order to test how the number of different models for a given category (e.g. number of models with SASI, or with a neutrino-driven shock revival mechanism) influences the accuracy of our algorithms, we tested the performance of the codes by leaving out one of the models from the training sample entirely and then running the prediction step only on the left-out model. This analysis has been performed 14 times, leaving out a different SN model each time. The results clearly show that if we detect a signal which originates from a SN different from all of the models used in the training sample, the predictions will only be viable if we have enough models near the unknown waveform in the parameter space. Of all of the categories we studied, the one having the largest number of models is the class of neutrino-driven explosions. Figure \ref{fig:leftout} shows the balanced accuracy for the predicted shock revival mechanism for this class of models. From all of the combinations tried, PCA+KNN is the most accurate, and PCA+SVM performs the least accurately. For the rest of the model classes, we did not have enough different models to cover the parameter space sufficiently, so we did not achieve a balanced accuracy above $0.5$.

\section{Conclusion} \label{sec:conclusion}

Many previous papers have already discussed the detection efficiency of the LIGO-Virgo-KAGRA network for various core-collapse supernova (CCSN) signals, and in this paper we took a step forward to explore the potential of GW observations to shed light on various aspects of the CCSN process. Using 14 advanced 3D hydrodynamic simulations, we generated GW waveforms and analyzed them using the LVK network during the O5 run. We then employed the SN-optimized BayesWave to reconstruct the waveforms and used these reconstructions to train and evaluate various machine learning techniques. The goal was to determine if the shock revival process, key features, or physical parameters of CCSNe could be inferred from the observation of gravitational waves.

We studied four different classification problems, which included the determination of the shock revival mechanism, and the presence of SASI, prompt convection and rotation. For nearly all cases, we got a balanced accuracy above 90\% for SNR $\gtrsim$ 30, and generally a higher SNR led to better classification. It is not straightforward to translate the SNR values to distances as various CCSN models have widely different energies. However, we generally do not expect detections with such high SNRs from neutrino-driven SN explosions farther away than a few kpcs, and from magneto-rotational SNe farther away than a few tens of kpcs. Beside the classifications, we also used our techniques to infer the progenitor masses and the angular velocities of the rotation. Generally the progenitor mass was more accurately estimated than the angular velocity. Even for high SNRs, the normalized root mean squared errors of the masses stayed above $\sim30\%$ and those of the angular velocity above $\sim 60\%$. As we reach a limit for high SNRs, the precision of the estimations cannot be increased by the improvement of the GW detectors, but only by performing more CCSN simulations, maybe even with the explicit intention of trying to cover the parameter space uniformly. 

Finally, we ran the prediction phase exclusively on the model that was completely omitted out of the training sample, testing how well the codes performed in this case. There have been 14 iterations of this analysis, omitting a different SN model each time. The findings clearly demonstrate that the predictions will only be accurate if there are sufficient models in the parameter space close to the unknown waveform. Hence, if we are going to detect a high SNR CCSN with a waveform similar to one of the models we have used, our pipeline would be able to determine the explosion mechanism and the presence of SASI, prompt convection and rotation from the GW waveform, however in the event that we detect a signal coming from an SN that is distinct from all of the models used in the training sample, the accuracy will be lower. Therefore, as newer and newer 3D CCSN simulations become available, we plan to incorporate the ones distinctly different from these 14 models to our pipeline, so that we can maximize the scientific reward of the first CCSN GW detection.

\section*{Acknowledgements}
This paper was reviewed by the LIGO Scientific Collaboration (document number: P2300045) and by the Virgo Collaboration (document number: VIR-0148A-23). The authors would like to thank Freija Beirnaert, Archisman Ghosh, Peter Raffai, Daniela Pascucci and Cezary Turski for fruitful discussions throughout the project. We are thankful for the feedback of the BayesWave development team. The authors thank Maria Lisa Brozzetti, Giuseppe Greco and Jade Powell for their useful comments on the manuscript. GD is supported through the iBOF-project BOF20/IBF/124. 
TS is supported by the J\'anos Bolyai Research Scholarship of the Hungarian Academy of Sciences, as well as by the FK134432 grant of the National Research, Development and Innovation (NRDI) Office of Hungary and the \'UNKP 22-5 New National Excellence Programs of the Ministry for Culture and Innovation from the source of the NRDI Fund, Hungary. The authors are grateful for computational resources provided by the LIGO Lab and supported by the National Science Foundation Grants PHY-0757058 and PHY-0823459. 
We acknowledge the use of the following packages in this analysis: Astropy \citep{2013A&A...558A..33A, 2018AJ....156..123A}, Matplotlib \citep{Hunter:2007}, NumPy \citep{2011CSE....13b..22V}, Pandas \citep{Pandas}, Scikit-learn \citep{scikit-learn}, SciPy \citep{scipy}, Seaborn \citep{Seaborn}, and UMAP \citep{McInnes2018,2018arXiv180203426M}.\\

\bibliography{SN_paper}{}

\begin{thebibliography}{}
\expandafter\ifx\csname natexlab\endcsname\relax\def\natexlab#1{#1}\fi
\providecommand{\url}[1]{\href{#1}{#1}}
\providecommand{\dodoi}[1]{doi:~\href{http://doi.org/#1}{\nolinkurl{#1}}}
\providecommand{\doeprint}[1]{\href{http://ascl.net/#1}{\nolinkurl{http://ascl.net/#1}}}
\providecommand{\doarXiv}[1]{\href{https://arxiv.org/abs/#1}{\nolinkurl{https://arxiv.org/abs/#1}}}

\bibitem[{Aasi {et~al.}(2015)Aasi, Abbott, Abbott, Abbott, Abernathy, Ackley,
  Adams, Adams, Addesso, \& et~al.}]{Aasietal2015}
Aasi, J., Abbott, B.~P., Abbott, R., {et~al.} 2015, Classical and Quantum
  Gravity, 32, 074001, \dodoi{10.1088/0264-9381/32/7/074001}

\bibitem[{{Abbott} {et~al.}(2018){Abbott}, {Abbott}, {Abbott},
  {et~al.}}]{2018LRR....21....3A}
{Abbott}, B.~P., {Abbott}, R., {Abbott}, T.~D., {et~al.} 2018, Living Reviews
  in Relativity, 21, 3, \dodoi{10.1007/s41114-018-0012-9}

\bibitem[{{Abbott} {et~al.}(2020){Abbott}, {Abbott}, {Abbott},
  {et~al.}}]{2020PhRvD.101h4002A}
---. 2020, \prd, 101, 084002, \dodoi{10.1103/PhysRevD.101.084002}

\bibitem[{{Abbott} {et~al.}(2021{\natexlab{a}}){Abbott}, {Abbott}, {Acernese},
  {et~al.}}]{2021arXiv210801045T}
{Abbott}, R., {Abbott}, T.~D., {Acernese}, F., {et~al.} 2021{\natexlab{a}},
  arXiv e-prints, arXiv:2108.01045, \dodoi{10.48550/arXiv.2108.01045}

\bibitem[{{Abbott} {et~al.}(2021{\natexlab{b}}){Abbott}, {Abbott}, {Acernese},
  {et~al.}}]{2021arXiv211103606T}
---. 2021{\natexlab{b}}, arXiv e-prints, arXiv:2111.03606.
\newblock \doarXiv{2111.03606}

\bibitem[{{Abbott} {et~al.}(2021{\natexlab{c}}){Abbott}, {Abbott}, {Acernese},
  {et~al.}}]{2021PhRvD.104l2004A}
---. 2021{\natexlab{c}}, \prd, 104, 122004, \dodoi{10.1103/PhysRevD.104.122004}

\bibitem[{{Abdikamalov} {et~al.}(2014){Abdikamalov}, {Gossan}, {DeMaio}, \&
  {Ott}}]{2014PhRvD..90d4001A}
{Abdikamalov}, E., {Gossan}, S., {DeMaio}, A.~M., \& {Ott}, C.~D. 2014, \prd,
  90, 044001, \dodoi{10.1103/PhysRevD.90.044001}

\bibitem[{Acernese {et~al.}(2014)Acernese, Agathos, Agatsuma, Aisa, Allemandou,
  Allocca, Amarni, Astone, Balestri, Ballardin, \& et~al.}]{Acernese_2014}
Acernese, F., Agathos, M., Agatsuma, K., {et~al.} 2014, Classical and Quantum
  Gravity, 32, 024001, \dodoi{10.1088/0264-9381/32/2/024001}

\bibitem[{Akutsu {et~al.}(2020)Akutsu, Ando, Arai,
  {et~al.}}]{akutsu2020overview}
Akutsu, T., Ando, M., Arai, K., {et~al.} 2020, Overview of KAGRA: Detector
  design and construction history, \dodoi{10.1103/PhysRevD.101.084002}

\bibitem[{Andresen {et~al.}(2017)Andresen, Müller, Müller, \&
  Janka}]{10.1093/mnras/stx618}
Andresen, H., Müller, B., Müller, E., \& Janka, H.-T. 2017, Monthly Notices
  of the Royal Astronomical Society, 468, 2032, \dodoi{10.1093/mnras/stx618}

\bibitem[{Andresen {et~al.}(2019)Andresen, Müller, Janka, Summa, Gill, \&
  Zanolin}]{10.1093/mnras/stz990}
Andresen, H., Müller, E., Janka, H.-T., {et~al.} 2019, Monthly Notices of the
  Royal Astronomical Society, 486, 2238, \dodoi{10.1093/mnras/stz990}

\bibitem[{{Antelis} {et~al.}(2022){Antelis}, {Cavaglia}, {Hansen},
  {et~al.}}]{2022PhRvD.105h4054A}
{Antelis}, J.~M., {Cavaglia}, M., {Hansen}, T., {et~al.} 2022, \prd, 105,
  084054, \dodoi{10.1103/PhysRevD.105.084054}

\bibitem[{{Astropy Collaboration} {et~al.}(2013){Astropy Collaboration},
  {Robitaille}, {Tollerud}, {Greenfield}, {Droettboom}, {Bray}, {Aldcroft},
  {Davis}, {Ginsburg}, {Price-Whelan}, {Kerzendorf}, {Conley}, {Crighton},
  {Barbary}, {Muna}, {Ferguson}, {Grollier}, {Parikh}, {Nair}, {Unther},
  {Deil}, {Woillez}, {Conseil}, {Kramer}, {Turner}, {Singer}, {Fox}, {Weaver},
  {Zabalza}, {Edwards}, {Azalee Bostroem}, {Burke}, {Casey}, {Crawford},
  {Dencheva}, {Ely}, {Jenness}, {Labrie}, {Lim}, {Pierfederici}, {Pontzen},
  {Ptak}, {Refsdal}, {Servillat}, \& {Streicher}}]{2013A&A...558A..33A}
{Astropy Collaboration}, {Robitaille}, T.~P., {Tollerud}, E.~J., {et~al.} 2013,
  \aap, 558, A33, \dodoi{10.1051/0004-6361/201322068}

\bibitem[{{Astropy Collaboration} {et~al.}(2018){Astropy Collaboration},
  {Price-Whelan}, {Sip{\H o}cz}, {G{\"u}nther}, {Lim}, {Crawford}, {Conseil},
  {Shupe}, {Craig}, {Dencheva}, {Ginsburg}, {VanderPlas}, {Bradley},
  {P{\'e}rez-Su{\'a}rez}, {de Val-Borro}, {Aldcroft}, {Cruz}, {Robitaille},
  {Tollerud}, {Ardelean}, {Babej}, {Bach}, {Bachetti}, {Bakanov}, {Bamford},
  {Barentsen}, {Barmby}, {Baumbach}, {Berry}, {Biscani}, {Boquien}, {Bostroem},
  {Bouma}, {Brammer}, {Bray}, {Breytenbach}, {Buddelmeijer}, {Burke},
  {Calderone}, {Cano Rodr{\'{\i}}guez}, {Cara}, {Cardoso}, {Cheedella},
  {Copin}, {Corrales}, {Crichton}, {D'Avella}, {Deil}, {Depagne}, {Dietrich},
  {Donath}, {Droettboom}, {Earl}, {Erben}, {Fabbro}, {Ferreira}, {Finethy},
  {Fox}, {Garrison}, {Gibbons}, {Goldstein}, {Gommers}, {Greco}, {Greenfield},
  {Groener}, {Grollier}, {Hagen}, {Hirst}, {Homeier}, {Horton}, {Hosseinzadeh},
  {Hu}, {Hunkeler}, {Ivezi{\'c}}, {Jain}, {Jenness}, {Kanarek}, {Kendrew},
  {Kern}, {Kerzendorf}, {Khvalko}, {King}, {Kirkby}, {Kulkarni}, {Kumar},
  {Lee}, {Lenz}, {Littlefair}, {Ma}, {Macleod}, {Mastropietro}, {McCully},
  {Montagnac}, {Morris}, {Mueller}, {Mumford}, {Muna}, {Murphy}, {Nelson},
  {Nguyen}, {Ninan}, {N{\"o}the}, {Ogaz}, {Oh}, {Parejko}, {Parley}, {Pascual},
  {Patil}, {Patil}, {Plunkett}, {Prochaska}, {Rastogi}, {Reddy Janga},
  {Sabater}, {Sakurikar}, {Seifert}, {Sherbert}, {Sherwood-Taylor}, {Shih},
  {Sick}, {Silbiger}, {Singanamalla}, {Singer}, {Sladen}, {Sooley},
  {Sornarajah}, {Streicher}, {Teuben}, {Thomas}, {Tremblay}, {Turner},
  {Terr{\'o}n}, {van Kerkwijk}, {de la Vega}, {Watkins}, {Weaver}, {Whitmore},
  {Woillez}, {Zabalza}, \& {Astropy Contributors}}]{2018AJ....156..123A}
{Astropy Collaboration}, {Price-Whelan}, A.~M., {Sip{\H o}cz}, B.~M., {et~al.}
  2018, \aj, 156, 123, \dodoi{10.3847/1538-3881/aabc4f}

\bibitem[{{B{\'e}csy} {et~al.}(2017){B{\'e}csy}, {Raffai}, {Cornish}, {Essick},
  {Kanner}, {Katsavounidis}, {Littenberg}, {Millhouse}, \&
  {Vitale}}]{BayesWavePE}
{B{\'e}csy}, B., {Raffai}, P., {Cornish}, N.~J., {et~al.} 2017, \apj, 839, 15,
  \dodoi{10.3847/1538-4357/aa63ef}

\bibitem[{{Bethe} \& {Wilson}(1985)}]{1985ApJ...295...14B}
{Bethe}, H.~A., \& {Wilson}, J.~R. 1985, \apj, 295, 14, \dodoi{10.1086/163343}

\bibitem[{{Bisnovatyi-Kogan}(1971)}]{1971SvA....14..652B}
{Bisnovatyi-Kogan}, G.~S. 1971, \sovast, 14, 652

\bibitem[{{Bizouard} {et~al.}(2021){Bizouard}, {Maturana-Russel},
  {Torres-Forn{\'e}}, {Obergaulinger}, {Cerd{\'a}-Dur{\'a}n}, {Christensen},
  {Font}, \& {Meyer}}]{2021PhRvD.103f3006B}
{Bizouard}, M.-A., {Maturana-Russel}, P., {Torres-Forn{\'e}}, A., {et~al.}
  2021, \prd, 103, 063006, \dodoi{10.1103/PhysRevD.103.063006}

\bibitem[{Blondin {et~al.}(2003)Blondin, Mezzacappa, \&
  DeMarino}]{Blondin_2003}
Blondin, J.~M., Mezzacappa, A., \& DeMarino, C. 2003, The Astrophysical
  Journal, 584, 971, \dodoi{10.1086/345812}

\bibitem[{{Bruel} {et~al.}(2023){Bruel}, {Bizouard}, {Obergaulinger},
  {Maturana-Russel}, {Torres-Forn{\'e}}, {Cerd{\'a}-Dur{\'a}n}, {Christensen},
  {Font}, \& {Meyer}}]{2023arXiv230110019B}
{Bruel}, T., {Bizouard}, M.-A., {Obergaulinger}, M., {et~al.} 2023, arXiv
  e-prints, arXiv:2301.10019, \dodoi{10.48550/arXiv.2301.10019}

\bibitem[{{Bugli} {et~al.}(2022){Bugli}, {Guilet}, {Foglizzo}, \&
  {Obergaulinger}}]{2022arXiv221005012B}
{Bugli}, M., {Guilet}, J., {Foglizzo}, T., \& {Obergaulinger}, M. 2022, arXiv
  e-prints, arXiv:2210.05012.
\newblock \doarXiv{2210.05012}

\bibitem[{Bugli {et~al.}(2021)Bugli, Guilet, \&
  Obergaulinger}]{10.1093/mnras/stab2161}
Bugli, M., Guilet, J., \& Obergaulinger, M. 2021, Monthly Notices of the Royal
  Astronomical Society, 507, 443, \dodoi{10.1093/mnras/stab2161}

\bibitem[{{Burrows} \& {Vartanyan}(2021)}]{2021Natur.589...29B}
{Burrows}, A., \& {Vartanyan}, D. 2021, \nat, 589, 29,
  \dodoi{10.1038/s41586-020-03059-w}

\bibitem[{{Cornish} \& {Littenberg}(2015)}]{BayesWaveII}
{Cornish}, N.~J., \& {Littenberg}, T.~B. 2015, Classical and Quantum Gravity,
  32, 135012, \dodoi{10.1088/0264-9381/32/13/135012}

\bibitem[{{Cornish} {et~al.}(2021){Cornish}, {Littenberg}, {B{\'e}csy},
  {Chatziioannou}, {Clark}, {Ghonge}, \& {Millhouse}}]{BayesWaveIII}
{Cornish}, N.~J., {Littenberg}, T.~B., {B{\'e}csy}, B., {et~al.} 2021, \prd,
  103, 044006, \dodoi{10.1103/PhysRevD.103.044006}

\bibitem[{{D{\'a}lya} {et~al.}(2021){D{\'a}lya}, {Raffai}, \&
  {B{\'e}csy}}]{BayesWaveEcc}
{D{\'a}lya}, G., {Raffai}, P., \& {B{\'e}csy}, B. 2021, Classical and Quantum
  Gravity, 38, 065002, \dodoi{10.1088/1361-6382/abd7bf}

\bibitem[{{Engels} {et~al.}(2014){Engels}, {Frey}, \&
  {Ott}}]{2014PhRvD..90l4026E}
{Engels}, W.~J., {Frey}, R., \& {Ott}, C.~D. 2014, \prd, 90, 124026,
  \dodoi{10.1103/PhysRevD.90.124026}

\bibitem[{{Gill} {et~al.}(2018){Gill}, {Wang}, {Valdez}, {Szczepanczyk},
  {Zanolin}, \& {Mukherjee}}]{2018arXiv180207255G}
{Gill}, K., {Wang}, W., {Valdez}, O., {et~al.} 2018, arXiv e-prints,
  arXiv:1802.07255, \dodoi{10.48550/arXiv.1802.07255}

\bibitem[{{Hirata} {et~al.}(1987){Hirata}, {Kajita}, {Koshiba}, {Nakahata},
  {Oyama}, {Sato}, {Suzuki}, {Takita}, {Totsuka}, {Kifune}, {Suda},
  {Takahashi}, {Tanimori}, {Miyano}, {Yamada}, {Beier}, {Feldscher}, {Kim},
  {Mann}, {Newcomer}, {van}, {Zhang}, \& {Cortez}}]{1987PhRvL..58.1490H}
{Hirata}, K., {Kajita}, T., {Koshiba}, M., {et~al.} 1987, \prl, 58, 1490,
  \dodoi{10.1103/PhysRevLett.58.1490}

\bibitem[{Hunter(2007)}]{Hunter:2007}
Hunter, J.~D. 2007, Computing In Science \& Engineering, 9, 90,
  \dodoi{10.1109/MCSE.2007.55}

\bibitem[{{Iess} {et~al.}(2023){Iess}, {Cuoco}, {Morawski}, {Nicolaou}, \&
  {Lahav}}]{2023A&A...669A..42I}
{Iess}, A., {Cuoco}, E., {Morawski}, F., {Nicolaou}, C., \& {Lahav}, O. 2023,
  \aap, 669, A42, \dodoi{10.1051/0004-6361/202142525}

\bibitem[{{Janka}(2012)}]{2012ARNPS..62..407J}
{Janka}, H.-T. 2012, Annual Review of Nuclear and Particle Science, 62, 407,
  \dodoi{10.1146/annurev-nucl-102711-094901}

\bibitem[{Janka(2017)}]{Janka2017}
Janka, H.-T. 2017, Neutrino-Driven Explosions (Cham: Springer International
  Publishing), 1095--1150, \dodoi{10.1007/978-3-319-21846-5_109}

\bibitem[{{Janka} \& {Mueller}(1996)}]{1996A&A...306..167J}
{Janka}, H.~T., \& {Mueller}, E. 1996, \aap, 306, 167

\bibitem[{Jones {et~al.}(2001)Jones, Oliphant, Peterson, {et~al.}}]{scipy}
Jones, E., Oliphant, T., Peterson, P., {et~al.} 2001, {SciPy}: Open source
  scientific tools for {Python}.
\newblock \url{http://www.scipy.org/}

\bibitem[{{Kistler} {et~al.}(2013){Kistler}, {Haxton}, \&
  {Y{\"u}ksel}}]{2013ApJ...778...81K}
{Kistler}, M.~D., {Haxton}, W.~C., \& {Y{\"u}ksel}, H. 2013, \apj, 778, 81,
  \dodoi{10.1088/0004-637X/778/1/81}

\bibitem[{{Kuroda} {et~al.}(2017){Kuroda}, {Kotake}, {Hayama}, \&
  {Takiwaki}}]{2017ApJ...851...62K}
{Kuroda}, T., {Kotake}, K., {Hayama}, K., \& {Takiwaki}, T. 2017, \apj, 851,
  62, \dodoi{10.3847/1538-4357/aa988d}

\bibitem[{{Kuroda} {et~al.}(2016){Kuroda}, {Kotake}, \&
  {Takiwaki}}]{2016ApJ...829L..14K}
{Kuroda}, T., {Kotake}, K., \& {Takiwaki}, T. 2016, \apjl, 829, L14,
  \dodoi{10.3847/2041-8205/829/1/L14}

\bibitem[{Li {et~al.}(2011)Li, Leaman, Chornock, Filippenko, Poznanski,
  Ganeshalingam, Wang, Modjaz, Jha, Foley, \& Smith}]{li11}
Li, W., Leaman, J., Chornock, R., {et~al.} 2011, Monthly Notices of the Royal
  Astronomical Society, 412, 1441, \dodoi{10.1111/j.1365-2966.2011.18160.x}

\bibitem[{{Littenberg} \& {Cornish}(2015)}]{BayesWaveI}
{Littenberg}, T.~B., \& {Cornish}, N.~J. 2015, \prd, 91, 084034,
  \dodoi{10.1103/PhysRevD.91.084034}

\bibitem[{{McInnes} {et~al.}(2018){McInnes}, {Healy}, \&
  {Melville}}]{2018arXiv180203426M}
{McInnes}, L., {Healy}, J., \& {Melville}, J. 2018, arXiv e-prints,
  arXiv:1802.03426, \dodoi{10.48550/arXiv.1802.03426}

\bibitem[{McInnes {et~al.}(2018)McInnes, Healy, Saul, \&
  Großberger}]{McInnes2018}
McInnes, L., Healy, J., Saul, N., \& Großberger, L. 2018, Journal of Open
  Source Software, 3, 861, \dodoi{10.21105/joss.00861}

\bibitem[{Mckinney(2011)}]{Pandas}
Mckinney, W. 2011, Python High Performance Science Computer

\bibitem[{Mezzacappa {et~al.}(2020)Mezzacappa, Marronetti, Landfield, Lentz,
  Yakunin, Bruenn, Hix, Messer, Endeve, Blondin, \&
  Harris}]{PhysRevD.102.023027}
Mezzacappa, A., Marronetti, P., Landfield, R.~E., {et~al.} 2020, Phys. Rev. D,
  102, 023027, \dodoi{10.1103/PhysRevD.102.023027}

\bibitem[{{Mukherjee} {et~al.}(2021){Mukherjee}, {Nurbek}, \&
  {Valdez}}]{2021PhRvD.103j3008M}
{Mukherjee}, S., {Nurbek}, G., \& {Valdez}, O. 2021, \prd, 103, 103008,
  \dodoi{10.1103/PhysRevD.103.103008}

\bibitem[{{Obergaulinger} \& {Aloy}(2020)}]{2020MNRAS.492.4613O}
{Obergaulinger}, M., \& {Aloy}, M.~{\'A}. 2020, \mnras, 492, 4613,
  \dodoi{10.1093/mnras/staa096}

\bibitem[{{O'Connor} \& {Couch}(2018)}]{2018ApJ...865...81O}
{O'Connor}, E.~P., \& {Couch}, S.~M. 2018, \apj, 865, 81,
  \dodoi{10.3847/1538-4357/aadcf7}

\bibitem[{{Ott} {et~al.}(2013){Ott}, {Abdikamalov}, {M{\"o}sta}, {Haas},
  {Drasco}, {O'Connor}, {Reisswig}, {Meakin}, \&
  {Schnetter}}]{2013ApJ...768..115O}
{Ott}, C.~D., {Abdikamalov}, E., {M{\"o}sta}, P., {et~al.} 2013, \apj, 768,
  115, \dodoi{10.1088/0004-637X/768/2/115}

\bibitem[{{Pan} {et~al.}(2021){Pan}, {Liebend{\"o}rfer}, {Couch}, \&
  {Thielemann}}]{2021ApJ...914..140P}
{Pan}, K.-C., {Liebend{\"o}rfer}, M., {Couch}, S.~M., \& {Thielemann}, F.-K.
  2021, \apj, 914, 140, \dodoi{10.3847/1538-4357/abfb05}

\bibitem[{Pedregosa {et~al.}(2011)Pedregosa, Varoquaux, Gramfort, Michel,
  Thirion, Grisel, Blondel, Prettenhofer, Weiss, Dubourg, Vanderplas, Passos,
  Cournapeau, Brucher, Perrot, \& Duchesnay}]{scikit-learn}
Pedregosa, F., Varoquaux, G., Gramfort, A., {et~al.} 2011, Journal of Machine
  Learning Research, 12, 2825

\bibitem[{{Powell} \& {M{\"u}ller}(2022)}]{2022PhRvD.105f3018P}
{Powell}, J., \& {M{\"u}ller}, B. 2022, \prd, 105, 063018,
  \dodoi{10.1103/PhysRevD.105.063018}

\bibitem[{Powell \& Müller(2019)}]{10.1093/mnras/stz1304}
Powell, J., \& Müller, B. 2019, Monthly Notices of the Royal Astronomical
  Society, 487, 1178, \dodoi{10.1093/mnras/stz1304}

\bibitem[{Powell \& Müller(2020)}]{10.1093/mnras/staa1048}
---. 2020, Monthly Notices of the Royal Astronomical Society, 494, 4665,
  \dodoi{10.1093/mnras/staa1048}

\bibitem[{Powell {et~al.}(2021)Powell, Müller, \&
  Heger}]{10.1093/mnras/stab614}
Powell, J., Müller, B., \& Heger, A. 2021, Monthly Notices of the Royal
  Astronomical Society, 503, 2108, \dodoi{10.1093/mnras/stab614}

\bibitem[{{Radice} {et~al.}(2019){Radice}, {Morozova}, {Burrows}, {Vartanyan},
  \& {Nagakura}}]{2019ApJ...876L...9R}
{Radice}, D., {Morozova}, V., {Burrows}, A., {Vartanyan}, D., \& {Nagakura}, H.
  2019, \apjl, 876, L9, \dodoi{10.3847/2041-8213/ab191a}

\bibitem[{{Raza} {et~al.}(2022){Raza}, {McIver}, {D{\'a}lya}, \&
  {Raffai}}]{BayesWaveSN}
{Raza}, N., {McIver}, J., {D{\'a}lya}, G., \& {Raffai}, P. 2022, \prd, 106,
  063014, \dodoi{10.1103/PhysRevD.106.063014}

\bibitem[{{Roma} {et~al.}(2019){Roma}, {Powell}, {Heng}, \&
  {Frey}}]{2019PhRvD..99f3018R}
{Roma}, V., {Powell}, J., {Heng}, I.~S., \& {Frey}, R. 2019, \prd, 99, 063018,
  \dodoi{10.1103/PhysRevD.99.063018}

\bibitem[{{Scheidegger} {et~al.}(2010){Scheidegger}, {K{\"a}ppeli},
  {Whitehouse}, {Fischer}, \& {Liebend{\"o}rfer}}]{2010A&A...514A..51S}
{Scheidegger}, S., {K{\"a}ppeli}, R., {Whitehouse}, S.~C., {Fischer}, T., \&
  {Liebend{\"o}rfer}, M. 2010, \aap, 514, A51,
  \dodoi{10.1051/0004-6361/200913220}

\bibitem[{{Szczepa{\'n}czyk} {et~al.}(2021){Szczepa{\'n}czyk}, {Antelis},
  {Benjamin}, {Cavagli{\`a}}, {Gondek-Rosi{\'n}ska}, {Hansen}, {Klimenko},
  {Morales}, {Moreno}, {Mukherjee}, {Nurbek}, {Powell}, {Singh},
  {Sitmukhambetov}, {Szewczyk}, {Valdez}, {Vedovato}, {Westhouse}, {Zanolin},
  \& {Zheng}}]{2021PhRvD.104j2002S}
{Szczepa{\'n}czyk}, M.~J., {Antelis}, J.~M., {Benjamin}, M., {et~al.} 2021,
  \prd, 104, 102002, \dodoi{10.1103/PhysRevD.104.102002}

\bibitem[{Tenenbaum {et~al.}(2000)Tenenbaum, de~Silva, \& Langford}]{isomap}
Tenenbaum, J.~B., de~Silva, V., \& Langford, J.~C. 2000, Science, 290, 2319,
  \dodoi{10.1126/science.290.5500.2319}

\bibitem[{{Torres-Rivas} {et~al.}(2019){Torres-Rivas}, {Chatziioannou},
  {Bauswein}, \& {Clark}}]{BayesWaveNSPostMerger}
{Torres-Rivas}, A., {Chatziioannou}, K., {Bauswein}, A., \& {Clark}, J.~A.
  2019, \prd, 99, 044014, \dodoi{10.1103/PhysRevD.99.044014}

\bibitem[{{Tsang} {et~al.}(2020){Tsang}, {Ghosh}, {Samajdar}, {Chatziioannou},
  {Mastrogiovanni}, {Agathos}, \& {Van Den Broeck}}]{BayesWavePostMergerEcho}
{Tsang}, K.~W., {Ghosh}, A., {Samajdar}, A., {et~al.} 2020, \prd, 101, 064012,
  \dodoi{10.1103/PhysRevD.101.064012}

\bibitem[{van~der Maaten \& Hinton(2008)}]{tsne}
van~der Maaten, L., \& Hinton, G.~E. 2008, Journal of Machine Learning
  Research, 9, 2579

\bibitem[{{van der Walt} {et~al.}(2011){van der Walt}, {Colbert}, \&
  {Varoquaux}}]{2011CSE....13b..22V}
{van der Walt}, S., {Colbert}, S.~C., \& {Varoquaux}, G. 2011, Computing in
  Science and Engineering, 13, 22, \dodoi{10.1109/MCSE.2011.37}

\bibitem[{Waskom(2021)}]{Seaborn}
Waskom, M. 2021, Journal of Open Source Software, 6, 3021,
  \dodoi{10.21105/joss.03021}

\bibitem[{{Woosley} \& {Heger}(2006)}]{2006ApJ...637..914W}
{Woosley}, S.~E., \& {Heger}, A. 2006, \apj, 637, 914, \dodoi{10.1086/498500}

\end{thebibliography}
\bibliographystyle{aasjournal}

\end{document}